# Antigène HBs positif en l'absence de toute infection par le virus de l'hépatite B : la face cachée de l'Ag HBs

## Positive HBs antigen in the absence of hepatitis B virus infection

José Bras Cachinho [1], Maud François [2], Karl Stefic [1,3], Julien Marlet [1,3]

[1] Bactériologie-Virologie-Hygiène, CHU de Tours, France

[2] Hémodialyse, CHU de Tours, France

[3] Unité INSERM U1259, Université de Tours, France

**Résumé.** La stratégie de dépistage d'une infection par le virus de l'hépatite B (VHB) a évolué en 2019 avec la recherche systématique de l'Ag HBs, des Ac anti-HBs et des Ac anti-HBc. Ces trois marqueurs permettent d'identifier les patients infectés par le VHB, les patients vaccinés contre le VHB, ceux ayant eu un contact avec le VHB et ceux n'ayant jamais été en contact avec le VHB. Pour prévenir toute erreur d'interprétation, la conclusion de la sérologie VHB doit tenir compte du contexte clinico-biologique. Dans le cas particulier d'un Ag HBs positif, il faudra évoquer en première intention une infection par le VHB. En seconde intention, notamment en l'absence d'Ac anti-HBc, de cytolyse hépatique et d'ictère, il faudra évoquer la possibilité d'une vaccination VHB récente, d'un Ag HBs faux positif voire d'une infection débutante. L'interrogatoire permet de révéler une vaccination récente tandis qu'un Ag HBs faussement positif peut être détecté avec un test de confirmation de l'Ag HBs. Si nécessaire, un contrôle sérologique après 2 à 3 semaines peut être réalisé pour confirmer ou infirmer l'infection par le VHB.

**Mots clés** : dépistage, VHB, sérologie, Vidas, Architect, vaccination

**Abstract**. French recommendations for the screening of hepatitis B virus (HBV) infection were updated in 2019 with the association of three markers: HBs Ag, anti-HBs Ab and anti-HBc Ab. These three markers allow identification of infected patients, vaccinated patients and patients who have been in contact with HBV. A positive HBs Ag is usually associated with HBV infection but this interpretation must take into account the clinical context. In particular, the absence of anti-HBc Ab, normal ALAT levels and the absence of jaundice can be associated with recent HBV vaccination or false-positive HBs Ag. Recent HBV vaccination can usually be confirmed by patient questioning, while confirmatory tests are useful to detect false positive HBs Ag. If necessary, a second sample can be requested to confirm the interpretation.

**Manuscrit**

Les deux principales indications de la sérologie hépatite B sont le dépistage de l'infection par le virus de l'hépatite B (VHB) et le contrôle post-vaccinal. Ces indications regroupent un grand nombre de situations cliniques qu'il faut connaitre pour interpréter correctement un antigène HBs (Ag HBs) positif.

Le dépistage de l'infection par le VHB est obligatoire en début de grossesse et est recommandé chez les personnes ayant des comportements sexuels à risque, les partenaires de sujets porteurs chroniques, les sujets provenant de zones de forte endémie et les usagers de drogues par voie intraveineuse [1], [2]. La stratégie de dépistage recommandée par la Haute Autorité de Santé (HAS) et la nomenclature des actes de Biologie Médicale (NABM) a récemment a évoluée en 2019 et associe trois marqueurs sérologiques : Ag HBs, Ac anti-HBs et Ac anti-HBc (Figure 1) [1], [3]. Ces trois marqueurs permettent d'identifier spécifiquement les patients infectés par le VHB, les patients vaccinés contre le VHB, ceux ayant eu un contact avec le VHB et ceux n'ayant jamais été en contact avec le VHB. A noter que dans le cas particulier de la femme enceinte, le dépistage obligatoire repose uniquement sur la recherche de l'Ag HBs en début de grossesse [2]–[4]. Pour prévenir toute erreur d'interprétation, la conclusion de la sérologie VHB doit tenir compte du contexte clinique et du résultat des trois marqueurs (hors femme enceinte). En cas d'Ag HBs positif, un contrôle sur un second prélèvement n'est plus nécessaire mais une vigilance particulière est indispensable avant de conclure à une infection par le VHB. Nous illustrons ici à travers 3 cas de patients suivis au CHU de Tours, une situation clinique précise où l'Ag HBs est présent et détectable dans le sang du patient en l'absence d'infection par le VHB.

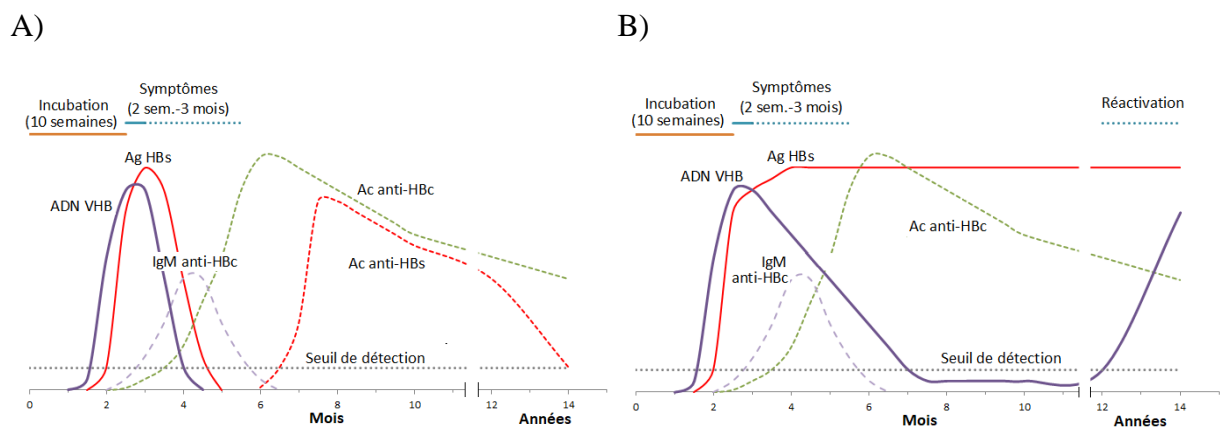

*Figure 1 : Cinétique des marqueurs sérologiques au cours A) d'une hépatite B aiguë et B) d'une hépatite B chronique.*

*L'Ag HBs, glycoprotéine d'enveloppe du VHB, est positif en cas d'infection par le VHB. Sa détection pendant plus de 6 mois définie l'hépatite B chronique. Les Ac anti-HBc, dirigés contre la capside du VHB, sont positifs en cas de contact avec le VHB. Les Ac anti-HBs sont positifs en cas de vaccination ou de contact avec le VHB. L'ADN du VHB se positive un mois environ après la contamination. Il est suivi de l'Ag HBs qui précède d'environ une semaine les symptômes (ictère, augmentation des transaminases) et l'apparition des IgM et Ac anti-HBc.*

*Les IgM anti-HBc disparaissent en 6 mois alors que les Ac anti-HBc restent positifs. Les Ac anti-HBs deviennent détectables uniquement après disparition de l'Ag HBs [5].*

**Cas clinico-biologiques**

Le premier patient était un homme de 61 ans hospitalisé dans le service de Néphrologie du CHU de Tours pour un bilan pré-dialyse dans un contexte de dégradation de la fonction rénale secondaire à une hypertension artérielle chronique familiale. Le bilan virologique réalisé retrouvait un Ag HBs positif (ratio signal patient/seuil = 7,25 ; seuil défini par calibration) avec le test Architect Ag HBs qualitatif (Abbott) sur le système automatisé Architect i2000SR. Sur ce même système, les tests Architect Ac anti-HBs (Abbott) et Architect Ac anti-HBc (Abbott) étaient négatifs. Le résultat de l'Ag HBs était discordant avec les antériorités (Ag HBs connu négatif dans le laboratoire) et avec le contexte clinique (absence de cytolyse hépatique et d'ictère). Dans ce contexte de découverte d'Ag HBs positif, la stratégie adoptée au CHU de Tours (hors recommandations) repose sur l'utilisation d'un second test (test VIDAS® Ag HBs Ultra, Biomérieux). L'Ag HBs a été confirmé positif avec ce second test (ratio patient/standard = 0,36 ; seuil de positivité = 0,13).

Le second patient était un homme de 85 ans, dialysé depuis 2 ans pour une insuffisance rénale chronique terminale sur néphropathie diabétique et vasculaire. Ce patient était connu avec un Ag HBs négatif, des Ac anti-HBs et des Ac anti-HBc négatifs. A l'occasion d'une consultation en néphrologie pour son bilan semestriel de dialyse, un Ag HBs positif a été détecté et confirmé en suivant une stratégie diagnostique similaire au premier patient (Tableau 1).

Le troisième patient était un nouveau-né hospitalisé en Réanimation Néonatale pour prématurité à 28 semaines d'aménorrhées et détresse respiratoire. Les sérologies effectuées chez la mère à l'accouchement avaient mis en évidence un Ag HBs négatif et des Ac anti-HBc négatifs. A l'âge de 7 semaines, dans un contexte d'accident d'exposition au sang, un Ag HBs positif a été détecté et confirmé en suivant une stratégie diagnostique similaire au premier patient (Tableau 1).

|  | Patient 1 | Patient 2 | Patient 3 |
|---|---|---|---|
| **Age** | 61 ans | 84 ans | 7 semaines |
| **Sexe** | Masculin | Masculin | Masculin |
| **Contexte** | Hémodialyse | Hémodialyse | Accident d'exposition au sang |
| **Sérologie initiale** |  |  |  |
| Architect Ag HBs qualitatif (ratio, seuil=1) | 7,25 | 4,01 | 156 |
| Vidas Ag HBs Ultra (ratio, seuil=0,13) | 0,36 | 0,26 | 7,09 |
| **Vaccination VHB antérieure** |  |  |  |
| Délai avant la sérologie (j) | 2 | 4 | 1 |
| Vaccin (dose d'Ag HBs) | Inconnu | Engerix B 20 µg (double dose=40 µg) | Infanrix Hexa (10 µg) |
| **Contrôle sérologique** |  |  |  |
| Délai après la sérologie initiale (j) | 173 | 39 | 14 |
| Architect Ag HBs qualitatif (ratio, seuil=1) | 0,20 | 0,24 | 0,20 |

| Architect Ac anti-HBs (UI/L, seuil=10 UI/L) | 33 | 0,1 | 77 |

*Tableau 1 : Détection d'un Ag HBs post-vaccination VHB chez trois patients et contrôle sérologique à distance montrant l'absence d'infection VHB.*

**SYNTHESE ET CONCLUSIONS CLINICO-BIOLOGIQUES**

Dans ces trois cas, la présence isolée d'Ag HBs sans Ac anti-HBc pouvait faire évoquer une infection débutante par le VHB. Toutefois, ces résultats étaient discordants avec les antériorités et avec le contexte clinique (patients connus Ag HBs négatif, absence de facteurs de risque, absence d'ictère et absence de cytolyse hépatique). L'ensemble de ces éléments doivent faire évoquer une vaccination récente contre le VHB avec détection de l'Ag HBs vaccinal. Cette hypothèse a été confirmée pour les trois patients après échanges avec le prescripteur. Un contrôle sérologique à distance a confirmé l'absence d'infection par le VHB avec Ac anti-HBc et Ag HBs négatifs. La vaccination a induit une réponse immunitaire satisfaisante chez les patients 1 et 3 mais pas chez le patient 2 (Tableau 1). A noter que ce patient était dialysé chronique, ce qui a pu contribuer à la mauvaise réponse vaccinale [6], [7].

**LE POINT DE VUE DU CLINICIEN**

La vaccination contre l'Hépatite B est devenue obligatoire pour les enfants nés à partir du premier janvier 2018. Le schéma vaccinal introduit à partir de 2013 préconise 3 injections à 2, 4 et 11 mois, par un vaccin combiné hexavalent contre la Diphtérie, le Tétanos, la Poliomyelite, la Coqueluche, les infections à *Haemophilus influenzae* de type b et l'Hépatite B [8]. La vaccination contre l'Hépatite B est également obligatoire chez les professionnels de santé exposés au virus et chez tout nouveau-né de mère porteuse de l'Ag HBs, associée dans ce cas avec des immunoglobulines spécifiques anti-HBs (sérovaccination). Le vaccin VHB est recommandé pour d'autres catégories de patients à risque, dont les insuffisants rénaux chroniques dialysés [8] et les patients ayant des Ac anti-HBc positifs isolés [4].

Il est recommandé de réaliser une sérologie VHB post-vaccinale au moins annuellement, chez les patients dialysés ou immunodéprimés, avec un rappel vaccinal si le taux d'Ac anti-HBs est inférieur à 10 UI/L [8]. Ce contrôle sérologique doit être réalisé à distance (2-3 semaines) de toute vaccination contre le VHB pour pouvoir interpréter le taux d'Ac anti-HBs et pour éviter toute détection d'Ag HBs vaccinal. Chez ces patients, des Ac anti-HBs <10 UI/L et/ou la détection d'Ac anti-HBc doivent faire rechercher un Ag HBs pour écarter toute infection ou réactivation du VHB [4].

## LE POINT DE VUE DU BIOLOGISTE

La détection d'Ag HBs vaccinaux s'explique par la composition des vaccins VHB actuellement commercialisés (Infanrix Hexa, Engerix B, HBVaxPro). En effet, ces derniers contiennent de l'Ag HBs recombinant non glycosylé produit en levures par génie génétique. Cet Ag HBs recombinant est non infectieux et est capable d'induire une réponse immunitaire neutralisante. Il peut être détecté, dans les jours suivants une vaccination contre le VHB, par les tests sérologiques ciblant l'Ag HBs en tant que glycoprotéine d'enveloppe du VHB [9]–[12]. Pour cette raison, un Ag HBs positif doit faire évoquer une infection en cours par le VHB ou, selon le contexte, un Ag HBs vaccinal.

Pour interpréter un résultat biologique, et notamment un Ag HBs positif, le biologiste doit tenir compte du risque de faux positifs (exceptionnels pour l'Ag HBs [4]) et du contexte clinico-biologique. Pour limiter le risque de faux positifs en Ag HBs, la stratégie adoptée au CHU de Tours (hors recommandations) repose sur l'utilisation d'un second test (test VIDAS® Ag HBs Ultra, Biomérieux) pour tout Ag HBs découvert positif en test Architect Ag HBs qualitatif. Entre février 2010 et mai 2018, 76 280 Ag HBs ont été testés négatifs et 2119 ont été testés positifs avec le test Architect Ag HBs qualitatif. Parmi ces 2119 Ag HBs positifs, 370 étaient des découvertes et ont donc été testés pour confirmation avec le test VIDAS® Ag HBs Ultra (Biomérieux). Ce test était positif pour 257 échantillons et négatif pour les 113 autres. Ces 113 échantillons ont été considérés comme de très probables faux positifs du test Architect sur la base du contexte clinique et d'au moins un facteur biologique parmi l'absence d'Ac anti-HBc (n=103) et/ou un ratio Architect Ag HBs qualitatif <5 fois le seuil (n=92). Lorsqu'un contrôle sérologique à distance a été réalisé, il a permis de confirmer l'absence d'infection (7 patients sur 7 contrôlés en 2017). Le contraste entre la très bonne spécificité du test Architect Ag HBs qualitatif sur la période étudiée (76 280 Ag HBs négatifs / (76 280+113) soit 99,9% de spécificité) et la fréquence élevée de faux positifs parmi l'ensemble des découvertes d'Ag HBs positifs (113/370=31%) est un argument en faveur de l'utilisation d'un test de confirmation (hors recommandations).

Avec ou sans confirmation, l'interprétation doit tenir compte du contexte clinico-biologique, notamment pour détecter d'éventuels Ag HBs vaccinaux. En effet, sur les 257 découvertes d'Ag HBs confirmés positif, l'analyse du contexte clinico-biologique a mis en évidence 246 infections par le VHB et 11 patients non infectés par le VHB, dont 8 vaccinations récentes (soit 3,1% des découvertes d'Ag HBs positifs) et 3 faux positifs. Ces faux positifs, inexpliqués mais rares, avaient un signal Architect Ag HBs qualitatif faible (ratio signal/seuil < 4,5), sans contexte infectieux ni vaccinal et ont été contrôlés négatifs à distance. Concernant la détection d'Ag HBs vaccinaux, dans notre expérience, les principaux facteurs cliniques associés étaient l'absence d'hépatite clinique (ictère) ou biologique (ALAT normales, résultats non montrés) et la prescription chez des patients exposés à des rappels vaccinaux réguliers contre le VHB (patients insuffisants rénaux hémodialysés ou patients immunodéprimés) [8].

Le principal facteur biologique orientant vers un Ag HBs vaccinal était l'absence d'Ac anti-HBc ($p<0.0001$, Figure 2). Dans notre expérience, la détection d'Ag HBs sans Ac anti-HBc (n=15 patients) était associée à une vaccination VHB récente (n=8), une infection par le VHB chez un patient immunodéprimé (n=3) ou non (n=1), ou un Ag HBs faux positif inexpliqué (n=3). Ce profil pourrait aussi faire évoquer une infection débutante avant l'apparition des Ac anti-HBc.

Enfin, dans certains cas, l'intensité du signal de l'Ag HBs pouvait guider l'interprétation (Figure 2). Un rapport signal Ag HBs/seuil > 200 était en faveur d'une infection par le VHB, tandis que des valeurs plus faibles ne permettaient pas de différencier une infection par le VHB d'une vaccination (Figure 2). Une vigilance accrue est nécessaire pour l'interprétation de ces valeurs faibles d'Ag HBs, en particulier avec les tests Ag HBs « ultra-sensibles » (limite de détection à 0,005 UI/mL contre 0,05 UI/mL pour les tests décrits ici) [13].

Dans tous les cas, le contrôle sérologique après 2-3 semaines permet de mettre en évidence la négativation de l'Ag HBs et le plus souvent l'apparition d'Ac anti-HBs. Lorsqu'il n'est pas possible d'attendre le contrôle sérologique, la recherche d'ADN du VHB (si disponible, hors recommandations) est utile pour confirmer l'absence d'infection par le VHB ($p<0.0001$, résultats non montrés, Figure 2).

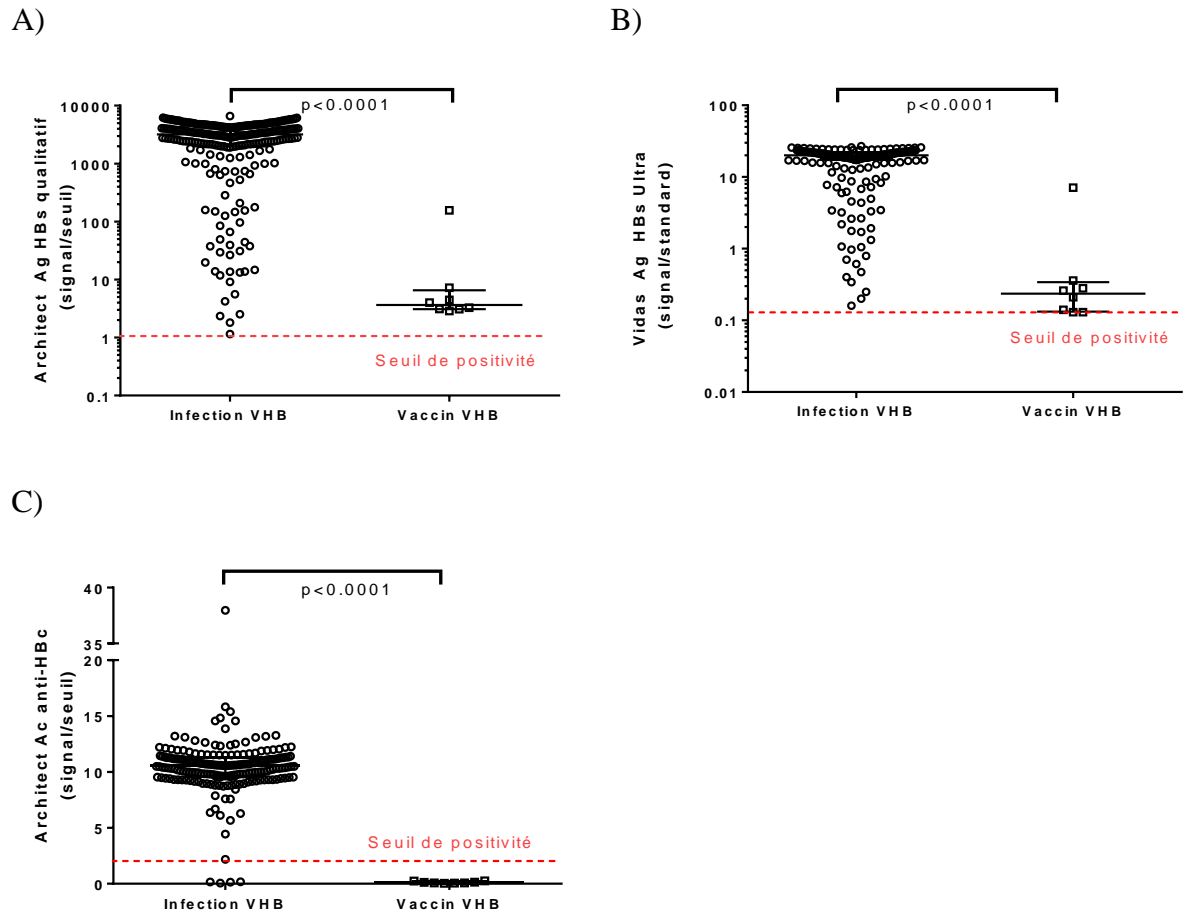

*Figure 2 : Profil sérologique du VHB associé aux 246 découvertes d'infection par le VHB et aux 8 patients en post-vaccination VHB. Sont représentés les médianes et les intervalles inter-quartiles. Les distributions ont été comparées en utilisant le test t de Student avec correction de Welch (C) ou le test de Mann-Whitney pour distributions non paramétriques (A et B), avec le logiciel GraphPad Prism v6.*

En conclusion, tout Ag HBs positif doit être interprété dans le contexte clinico-biologique et faire évoquer une infection par le VHB, sauf cas spécifique. En particulier, l'absence d'ictère, de cytolyse hépatique, d'Ac anti-HBc et un titre faible d'Ag HBs (selon les techniques) doivent faire suspecter un faux positif, une vaccination récente ou une infection débutante. Une vaccination VHB récente est détectable à l'interrogatoire et un Ag HBs faussement positif peut être détecté avec un test de confirmation de l'Ag HBs (hors recommandations). Si nécessaire, un contrôle sérologique après 2 à 3 semaines peut être réalisé (hors recommandations) pour confirmer ou infirmer l'infection par le VHB.

**BIBLIOGRAPHIE**